\documentclass[]{article}
\usepackage{packages}

\usepackage{arxiv}

\usepackage[utf8]{inputenc} 
\usepackage[T1]{fontenc}    
\usepackage{booktabs}       
\usepackage{amsfonts}       
\usepackage{nicefrac}       
\usepackage{microtype}      
\usepackage{doi}
\usepackage{natbib}

\title{Closed-loop Identification of a MSW Grate Incinerator using Bayesian Optimization for Selecting Model Inputs and Structure}

\date{December 20, 2023}	

\author{ Johannes Lips \\
	Institute of Combustion and Power Plant Technology\\
	University of Stuttgart\\
	Stuttgart, Germany \\
	\texttt{johannes.lips@ifk.uni-stuttgart.de} \\
        \And
        Stefan DeYoung \\
	MARTIN GmbH für Umwelt- und Energietechnik\\
	Munich, Germany \\
	\And
        Max Schönsteiner \\
	MARTIN GmbH für Umwelt- und Energietechnik\\
	Munich, Germany \\
	\And
	\href{https://orcid.org/0000-0002-0208-4100}{\includegraphics[scale=0.06]{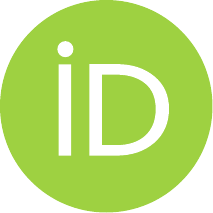}\hspace{1mm}Hendrik Lens} \\
	Institute of Combustion and Power Plant Technology\\
	University of Stuttgart\\
	Stuttgart, Germany \\
}

\renewcommand{\shorttitle}{Closed-loop Identification of a MSW Grate Incinerator using Bayesian Optimization}

\hypersetup{
pdftitle={Closed-loop Identification of a MSW Grate Incinerator using Bayesian Optimization for Selecting Model Inputs and Structure},
pdfauthor={J. Lips, S. DeYoung, M. Schoensteiner, H. Lens},
pdfkeywords={System identification, Municipal Solid Waste (MSW) incineration, Grate incineration, Dynamic process modeling, Bayesian Optimization},
}

\begin{document}
\maketitle

\begin{abstract}
The creation of low-order dynamic models for complex industrial systems is complicated by disturbances and limited sensor accuracy.
This work presents a system identification procedure that uses machine learning methods and process knowledge to robustly identify a low-order closed-loop model of a municipal solid waste (MSW) grate incineration plant.
These types of plants are known for their strong disturbances coming from fuel composition fluctuations.
Using Bayesian optimization, the algorithm ranks and selects inputs from the available sensor data and chooses the model structure.
This results in accurate models with low complexity while avoiding overfitting.
The method is applied and validated using data of an industrial MSW incineration plant.
The obtained models give excellent predictions and confidence intervals for the steam capacity and intermediate quantities such as supply air flow and flue gas temperature.
The identified continuous-time models are fully given, and their step-response dynamics are discussed.
The models can be used to develop model-based unit control schemes for grate incineration plants.
The presented method shows great potential for the identification of over-actuated systems or disturbed systems with many sensors.

\end{abstract}



\setcounter{footnote}{0} 
\section{Introduction, Motivation and Objective}
\subsection{Introduction to MSW Grate Incineration}
With the volume of generated municipal solid waste (MSW) globally increasing, MSW incineration offers a way to reduce the volume of waste, to thermally decompose its harmful and toxic compounds, and to recover energy released in the process as useful heat or electric power \citep{makarichiEvolutionWastetoenergyIncineration2018}.
Grate incineration is a particularly common and well-established method for MSW incineration, as it is able to process large amounts of MSW without strong preprocessing (sorting/shredding).
Although the technology behind MSW incinerators is similar, installations are often made to order and therefore vary in their specifications.

\begin{figure}[htbp]
    \fontsize{8}{10}\selectfont
    \centering
    \def\svgwidth{0.6\textwidth}
    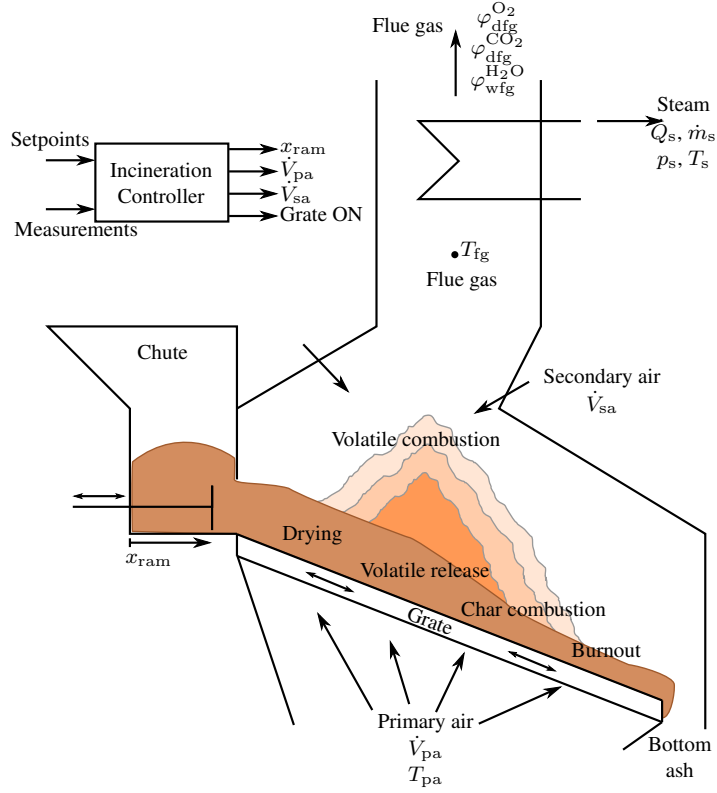
    \vspace{2pt}
    \caption{Grate incineration setup and control, main components and sensor measurements.}
    \label{fig:grateInc-overview}
\end{figure}    

A typical setup of a grate incineration plant, including the main components and the main measured variables, is given in \autoref{fig:grateInc-overview} and is described below.
The MSW is moved from a storage bunker to a feed hopper by a crane (not depicted).
From there, a ram feeder or stoker periodically pushes the waste onto the top of a moving grate. 
Depending on the layout of the plant, it is possible that there are multiple ram feeders and parallel grates.
As the waste travels downwards on the grate, the combustion process, consisting of multiple process steps, including drying, pyrolysis, gasification, as well as fixed carbon and volatiles combustion, takes place.
The main air supply, called primary air, is provided through the grate from below.
Secondary air, which is required to improve volatiles combustion, is supplied through nozzles located in the walls above the grate.
The steam generator, typically a drum boiler without reheating, cools the flue gas, which undergoes thorough flue gas cleaning before going through the stack.
In a connected steam cycle (not shown), the generated steam is used to produce electricity or heat for industrial processes and district heating \citep{qinDesignCombustionControl2008, leskensMIMOClosedloopIdentification2002}.

The high-level manipulated variables in the process are the movement of the ram feeder, $x_\str{ram}$, the volumetric flow rates of the primary and secondary air, $\VPair$ and $\VSair$, as well as the binary output `Grate ON', which is true when the grate is moving and false when not.
The speed at which the grate is moving and the distribution of the primary air under the grate could be added as additional manipulated variables.
However, in this work, these are assumed to be part of a lower-level control loop and are not considered.
The main controlled variables are (a) the live steam load ($\Qsteam$ in \si{MW} or as mass flow, $\mLS$, in \si{t/h}) (b) and the volumetric oxygen fraction at the end of the boiler $\Ot$, where the index dfg stands for \emph{dry flue gas}. 
The temperature of the flue gas in the region above the secondary air inlet $\Tfurn$ is also measured and controlled.
This is necessary, as the temperature must be kept above \qty{850}{\degC} for at least 2 seconds after the last secondary air inlet, in order to destroy dioxins and other unburnt flue gas components \citep{qinDesignCombustionControl2008}.
Other measured variables include the preheated primary air temperature $\TPair$, the volumetric carbon dioxide concentration at the boiler end $\COt$ and the volumetric water concentration at boiler end $\HtO$ (with wfg standing for \emph{wet flue gas}), as well as the live steam pressure $\pLS$ and temperature $T_\str{s}$.

Although grate incineration robustly processes different types of waste, the continuously varying composition, moisture content, as well as the heating value of the fuel and the varying time constants in the process result in a highly nonlinear system with strong disturbances, causing large fluctuations in the controlled variables.
Even in modern plants equipped with fast and accurate sensors and advanced MIMO control concepts, large deviations are common.
For example, peak deviations from steady-state setpoints from up to \qty{10}{\%} for the steam load and up to \qty{20}{\%} for the oxygen fraction setpoint were seen in the data analysed for this work (\qty{5}{s} resolution data).

The complexity of the process and the benefits associated with better control performance explain the persisting research in the field.
Many high-quality, complex mathematical, first-principle, and CFD dynamic models for both biomass and waste grate incineration can be found in the literature \citep{magnanelliDynamicModelingMunicipal2020, alobaidDynamicSimulationMunicipal2018, xiaTwofluidModelSimulation2020}.
These models are used to improve the understanding of the process and to develop new combustion control concepts with improved performance.

\subsection{Unit Control and Low-Order Models for MSW Grate Incineration}
Traditionally, MSW incineration plants set their operating point corresponding to the waste throughput or heating demand, making them baseload power plants that are inflexible from the power system perspective.
This mode of operation is not suited to the increasingly volatile production of renewable energy and the sector coupling that characterise modern energy systems.

A flexible operation method, either directly reacting to market price signals or as part of an optimized portfolio of generation units, would fit the needs of the energy system and increase possibilities for plant operators.
Practically, this would mean that a superordinate control system continuously controls the operating point of the plant with respect to price signals or coordinated together with the operating points of other portfolio units, using methods such as those outlined for thermal power stations in \citep{vdi-gesellschaftVDIVDE35082003}.
In order to use such model-based control approaches, low-order closed-loop plant models are needed.

The mentioned literature models do not qualify as model for this purpose.
These models are generally formulated as a set of differential-algebraic equations (DAEs), possibly containing partial differential equations.
However, model-based predictive control algorithms normally require a model that is based on ordinary differential equations (ODE).
Complex models also come with a high computational cost and typically take the (in reality unknown) fuel composition as input.
Linearization and model-order reduction (MOR) methods could be applied to find ODE formulations of the DAE models.
The highly individual character of the layout of MSW incineration plants would, however, require a complex model for each plant on which the superordinate control system is applied.
Even if linearization and MOR methods would result in useful low-order models, the modelling effort would therefore remain large for each new plant considered, when using a complex model as a starting point.

On the other hand, some research groups developed low-order models of grate incinerators that form a good basis for models suitable for unit control.
\citet{bauerModellingGrateCombustion2010} developed a low-order physics-based model for biomass grate incineration, which was successfully used in a model-predictive combustion controller \citep{kortelaFaulttolerantModelPredictive2015}.
This model takes the biomass inlet mass flow rate and moisture content as inputs.
This is possible for biomass with known density and moisture content, but not for MSW.

The models created by \citet{leskensMIMOClosedloopIdentification2002} and \citet{loosDynamicModelingCombustion1996} consider MSW as fuel and do not require the fuel composition as input.
Both used the MATLAB System Identification Toolbox \citep{ljungSystemIdentificationToolbox2023} to create models and obtained satisfactory results for the selected plant operating point.
However, in order to gather the data for model creation, excitation signals were added to the controller outputs, disturbing regular plant operation, which would ideally be avoided.
Also, the models did not attempt to capture any of the dynamics associated with the varying fuel composition, nor did they use additional available measurement signals to improve the model quality.
Furthermore, the possibilities of system identification have evolved since the development of these models.
Finally, these models only considered the main process behaviour and did not include the combustion controller.
A controller model would need to be added to these models to obtain a closed-loop model, as is required by a superordinate controller used for unit control.
It is not advisable to use the actual incineration controller implementation as part of the low-order model, because (a) the implementation might vary from plant to plant, risking an increased modelling effort, (b) the implementation might be highly nonlinear and complex, significantly increasing the complexity of the model, and (c) the implementation might not be available to the plant operator.

In summary, only few studies have looked into low-order modelling of MSW grate incineration, and both the physics-based and the system identification literature models are not readily applicable for unit control applications.

\subsection{Research Objective and Innovation}
The goal of this work is to develop a methodology with which low-order closed-loop MSW incineration models can be easily and robustly created for industrial MSW incineration plants using plant specifications and operational data.
Such models can serve as a basis for a model-based superordinate unit control, allowing MSW incineration plants to advance from their traditional inflexible operation to more flexible operation methods, in accordance with the needs of a modern energy system with high volatile renewable generation.
The models can also offer insight into the current plant dynamics and the effects of disturbances in different plants, without requiring extensive modelling effort.

To this end, we examine MSW grate incineration by combining physics-based and system identification approaches in a way that, to the best of our knowledge, has not been done in the literature before.
In order to achieve high model performance and obtain a robust method, we apply a new combination of modern expansions to traditional system identification, namely regularization, cross-validation, and hyperparameter tuning for model order and input selection through Bayesian optimization.
For easy usage in industrial plants, the method only takes signals that are available in a typical MSW plant as input.
The methodology is applied and validated using data from an industrial MSW incineration plant, and additional simulation responses with the identified models are presented to show the possibilities and limitations of the method.
The proposed system identification method is robust against the strong disturbances in the system.
Therefore, we believe it also has potential for system identification applications in other fields.

\subsection{Paper Structure}
This introduction is followed by a brief review of traditional system identification and an extensive discussion of the proposed new system identification procedure in \autoref{sec:sysID}.
This discussion is intentionally kept general (that is, not specific to the MSW incineration application) to allow readers to easily apply the method to other applications.
In the Modelling section (\autoref{sec:model}), the available plant measurement data is introduced, and its preprocessing is discussed before constructing two distinct models, both using the extended system identification procedure.
For each of the models, both the model setup and individual validation results are presented in dedicated subsections.
The full models are given in \autoref{sec:appFullModels}.
The models are compared and discussed in the Discussion section (\autoref{sec:discussion}).
That section also presents step responses illustrating the system's dynamic behaviour, which provide insight into the system and possible applications of the models.
A Final Remarks paragraph (\autoref{sec:final}) concludes the work.

\section{System Identification Procedure} \label{sec:sysID} 
\subsection{Traditional System Identification Procedure}
Following the notation introduced by \citet{ljungShiftParadigmSystem2020}, we consider a single experiment consisting of $N$ consecutive time domain observations of measurement data $(u,y)$, as well as a model $\mathcal{M}(\theta)$ with parameters $\theta$ taken from the model class 
\begin{equation}\label{eq:modelclass}
    \mathcal{M} = \left\{\mathcal{M}(\theta) \,|\,\theta \in D_\theta \right\} \punk{,}
\end{equation}
with $D_\theta$ the domain of the parameter space.
The output of the model, $\hat{y}(t|\theta)$, is obtained by applying the model $\mathcal{M}(\theta)$ to the input $u$.

For the given waste incineration process, a so-called process model with up to three real poles and a time delay,
\begin{equation} \label{eq:model-G}
    G_\vartheta (s) = 
    \frac{K_\str{p} \, e^{-sT_\str{d}}}
    {\prod_{i=1}^3 1+sT_{\str{p}i}}\,\Big|\, 
    \vartheta = \begin{bmatrix} K_\str{p} &\!\! T_\str{d} &\!\! T_{\str{p}1} &\!\! T_{\str{p}2} &\!\! T_{\str{p}3} \end{bmatrix} \punk{,}
\end{equation}
is used as general structure for model class $\mathcal{M}$.
For a system with $n$ inputs and a single output, the resulting model class is

\begin{equation} \label{eq:model-M}
    \mathcal{M} = \left\{\begin{bmatrix} G_{\vartheta_1} &\!\!\! G_{\vartheta_2} &\!\!\! ... &\!\!\! G_{\vartheta_{n}} \end{bmatrix}
    \,\Big|\, 
    \theta = \begin{bmatrix} \vartheta_1 &\!\!\! \vartheta_2 &\!\!\! ... &\!\!\! \vartheta_{n}\end{bmatrix} \in D_\theta
    \right\} \punk{.}
\end{equation}

We introduce the mean squared error, $\MSE$, as
\begin{equation}
\MSE\left(\theta\right)  =\frac{1}{N}\sum_{t=1}^N\left(y(t)-\hat{y}\left(t|\theta\right)\right)^2 \punk{.} \label{eq:MSE}
\end{equation} 
The coefficient-of-determination, $\Rt$, is a performance metric which is easy to interpret and can be related to the $\MSE$ as
\begin{equation}\label{eq:Rt}
    \Rt\left(\theta\right) = 1 - \frac{\MSE\left(\theta\right)}{\frac{1}{N}\sum_{t=1}^N\left(y(t)-\bar{y}\right)^2} \punk{.}
\end{equation}
with $\bar{y}$ the mean value of $y$ in the observed data.

In system identification, we are interested in the optimal parameters $\hat{\theta}$ and the corresponding optimal model $\mathcal{M}({\hat{\theta}})$ within the model class $\mathcal{M}$.
The optimum is expressed as
\begin{equation}\label{eq:thetaopt}
    \hat{\theta} = \argmin_{\theta \,\in\, D_\theta} \MSE(\theta)
\end{equation}
and can be found using the System Identification Toolbox in MATLAB \citep{ljungSystemIdentificationToolbox2023}.
To evaluate whether the identified model performs well in general, one can evaluate $\Rt$ and $\MSE$ on an unseen dataset.

\subsection{\sysIDHP{}: System Identification Procedure with Hyperparameter Tuning}

The complexity of the waste incineration system and high disturbances cause the traditional system identification procedure to fail, resulting in a high $\MSE$ on unseen data.
This failure can be due to
\begin{enumerate}
    \item \textit{underfitting} (bias error): $\mathcal{M}$ is too restrictive and does not contain any model that represents the true system with sufficient accuracy; 
    \item \textit{overfitting} (variance error): $\mathcal{M}({\hat{\theta}})$ was obtained by fitting not only the true system behaviour, but also the (stochastic) noise and disturbances, leading to higher $\MSE$ for validation data than for training data.
\end{enumerate}
Ideally, the model class is general enough to identify the true system dynamics accurately without additional degrees of freedom in $D_\theta$ that could be used to (falsely) identify noise and disturbances as part of the system dynamics.

A lot of research deals with the bias/variance trade-off problem in system identification.
Without trying to give a full overview, we mention some noteworthy studies and advancements that inspired this work.
Regularization with various kernels is often applied, for example by \citet{khosraviRegularizedSystemIdentification2020}.
\citet{tacxModelOrderSelection2021} have discussed model order selection in system identification, with a focus on robust control applications.
\citet{senaBayesianOptimizationNonlinear2021} have used Bayesian optimization to select the model order and the number of memory taps of a Volterra filter for the nonlinear system identification of optical transmitter filters. 
\citet{kivitsDynamicNetworkApproach2019} have presented an undirected graph-based approach as an alternative to the classical directional system identification.
This approach can help to combine physical system knowledge with system identification and to select the correct inputs for system identification in a classical sense.

The goal of this work is to deal with the bias/variance trade-off problem effectively for the described over-actuated disturbed MSW incineration system.
The resulting procedure should be easily applicable in industrial applications.
The procedure should also stay close to the traditional system identification that is known by practitioners and results in interpretable models.

\begin{figure*}[tbph]
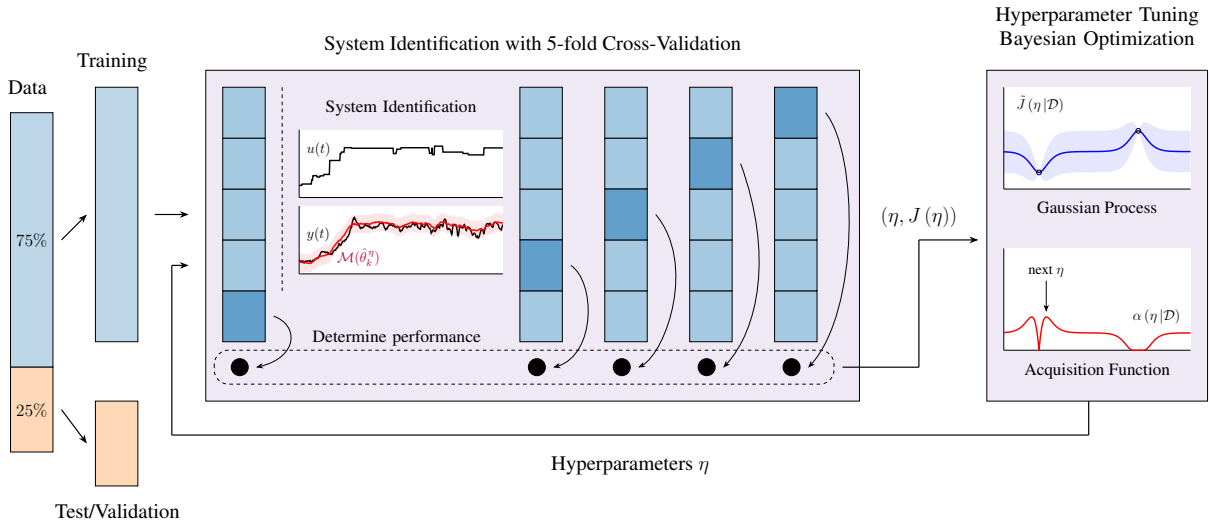

    \centering
    \resizebox{1\textwidth}{!}{
    \begin{tikzpicture}
\tikzstyle{every node}=[font=\LARGE]

\draw [ fill=pyth4!15 ] (33.75,6.5) rectangle (40.25,16.25);
\draw [ fill=pyth4!15 ] (10.75,6.5) rectangle (30,16.25);
\draw [ fill=pyth0!30 ] (5,15) rectangle  node {\Large $75\%$} (6.25,7.5);
\draw [ fill=pyth1!30 ] (5,7.5) rectangle  node {\Large $25\%$} (6.25,5);
\draw [ fill=pyth0!30 ] (7.5,15.75) rectangle (8.75,8.25);
\draw [ fill=pyth1!30 ] (7.5,6.5) rectangle (8.75,4);
\draw [line width=1pt, ->, >=Stealth] (6.5,11.25) -- (7.25,12);
\draw [line width=1pt, ->, >=Stealth] (6.5,6.25) -- (7.25,5.25);
\draw [ fill=pyth0!70 ] (11.25,8.25) rectangle (12.5,9.75);
\draw [ fill=pyth0!40 ] (11.25,9.75) rectangle (12.5,11.25);
\draw [ fill=pyth0!40 ] (11.25,11.25) rectangle (12.5,12.75);
\draw [ fill=pyth0!40 ] (11.25,12.75) rectangle (12.5,14.25);
\draw [ fill=pyth0!40 ] (11.25,14.25) rectangle (12.5,15.75);
\draw [dashed] (13,15.75) -- (13,9.75);
\node [font=\Large, anchor = north, text width = 6.8cm, align = center] at (16.5, 15.5) {System Identification};
\node [font=\Large, text width=6cm, align = center, anchor = south west] at (13.25,8) {Determine performance};
\draw [->, >=Stealth] (12.75,9) .. controls (13.5,8.75) and (13.5,7.75) .. (12.25,7.5);
\draw [ fill=black ] (11.75,7.5) circle (0.25cm);
\node [font=\LARGE] at (5.5,15.75) {Data};
\node [font=\LARGE] at (8,16.5) {Training};
\node [font=\LARGE] at (8.15,3.25) {Test/Validation};
\draw [ fill=pyth0!40 ] (20,8.25) rectangle (21.25,9.75);
\draw [ fill=pyth0!40 ] (20,11.25) rectangle (21.25,12.75);
\draw [ fill=pyth0!40 ] (20,12.75) rectangle (21.25,14.25);
\draw [ fill=pyth0!40 ] (20,14.25) rectangle (21.25,15.75);
\draw [ fill=pyth0!40 ] (22.5,8.25) rectangle (23.75,9.75);
\draw [ fill=pyth0!40 ] (22.5,9.75) rectangle (23.75,11.25);
\draw [ fill=pyth0!40 ] (22.5,12.75) rectangle (23.75,14.25);
\draw [ fill=pyth0!40 ] (22.5,14.25) rectangle (23.75,15.75);
\draw [ fill=pyth0!40 ] (25,8.25) rectangle (26.25,9.75);
\draw [ fill=pyth0!40 ] (25,9.75) rectangle (26.25,11.25);
\draw [ fill=pyth0!40 ] (25,11.25) rectangle (26.25,12.75);
\draw [ fill=pyth0!40 ] (25,14.25) rectangle (26.25,15.75);
\draw [ fill=pyth0!40 ] (27.5,8.25) rectangle (28.75,9.75);
\draw [ fill=pyth0!40 ] (27.5,9.75) rectangle (28.75,11.25);
\draw [ fill=pyth0!40 ] (27.5,11.25) rectangle (28.75,12.75);
\draw [ fill=pyth0!40 ] (27.5,12.75) rectangle (28.75,14.25);
\draw [ fill=pyth0!70 ] (20,9.75) rectangle (21.25,11.25);
\draw [ fill=pyth0!70 ] (22.5,11.25) rectangle (23.75,12.75);
\draw [ fill=pyth0!70 ] (25,12.75) rectangle (26.25,14.25);
\draw [ fill=pyth0!70 ] (27.5,14.25) rectangle (28.75,15.75);
\draw [ fill=black ] (20.5,7.5) circle (0.25cm);
\draw [ fill=black ] (23,7.5) circle (0.25cm);
\draw [ fill=black ] (25.5,7.5) circle (0.25cm);
\draw [ fill=black ] (28,7.5) circle (0.25cm);
\draw [->, >=Stealth] (21.5,10.5) .. controls (22.25,10) and (22.25,7.75) .. (21,7.5);
\draw [->, >=Stealth] (24,12) .. controls (25.25,10.5) and (24.25,7.75) .. (23.5,7.5);
\draw [->, >=Stealth] (26.5,13.5) .. controls (27.5,10.75) and (27,8.5) .. (26,7.5);
\draw [->, >=Stealth] (29,15) .. controls (30,13.75) and (30,9) .. (28.5,7.5);
\draw [rounded corners = 12.0, dashed] (11,8) rectangle  (29.25,7);
\draw [line width=1pt, ->, >=Stealth] (9.25,12) .. controls (9.75,12) and (9.75,12) .. (10.25,12);
\node [font=\LARGE] at (20.375,17) {System Identification with 5-fold Cross-Validation};
\draw [line width=1pt, ->, >=Stealth] (31.75,11.25) .. controls (32.75,11.25) and (32.75,11.25) .. (33.5,11.25);
\node [font=\LARGE, text width = 3.8cm, align = center, anchor = south] at (31.75,11.5) {$\left(\eta, J\left(\eta\right)\right)$};
\node [font=\LARGE, text width = 8cm,align = center] at (37,17.5) {Hyperparameter Tuning Bayesian Optimization};
\draw [line width=1pt, short] (36.75,6.5) .. controls (36.75,6) and (36.75,6) .. (36.75,5.5);
\draw [line width=1pt, short] (36.75,5.5) .. controls (23.25,5.5) and (23.25,5.5) .. (9.75,5.5);
\draw [line width=1pt, short] (9.75,5.5) .. controls (9.75,8) and (9.75,8) .. (9.75,10.5);
\draw [line width=1pt, ->, >=Stealth] (9.75,10.5) .. controls (10,10.5) and (10,10.5) .. (10.25,10.5);
\node [font=\LARGE, anchor = north] at (23.25,5) {Hyperparameters $\eta$};
\draw [line width=1pt, short] (29.5,7.5) .. controls (30.75,7.5) and (30.75,7.5) .. (31.75,7.5);
\draw [line width=1pt, short] (31.75,7.5) .. controls (31.75,9.5) and (31.75,9.5) .. (31.75,11.25);
\node [font=\Large,anchor=north] at (37,12.5) {Gaussian Process};
\node [font=\Large,anchor=north] at (37,7.75) {Acquisition Function};

\input{Figures/GP_n2_v1}
\node [font=\large, anchor = west] at (34.5,15.25) {$\tilde{J}\left(\eta\,|\mathcal{D}\right)$};

\input{Figures/acq_n2_v1}

\node [font=\large, anchor = south] at (35.5,10.05) {next $\eta$};
\draw [->, >=Stealth] (35.5,10.05) -- (35.5,9.15);
\node [font=\large, anchor = center] at (38.75,9.0) {$\alpha\left(\eta\,|\mathcal{D}\right)$};

\input{Figures/BasicModel_forprocessSketch}
\node [font=\large,anchor=north west] at (13.6,14.25) {$u(t)$};
\node [font=\large,anchor=north west] at (13.6,11.75) {$y(t)$};
\node [font=\large,anchor=west,color=purple] at (14.5,10.75) {$\mathcal{M}({\hat{\theta}_k^{\,\eta}})$};

\end{tikzpicture}
    }
    \caption{Overview of the system identification procedure with cross-validation and Bayesian optimization for hyperparameter tuning.}
    \label{fig:overview-sysIDHP}
\end{figure*}

The measures taken to improve the system identification performance are now formulated in terms of the changes in over- and underfitting aimed for.
The hyperparameters (tuneable parameters which are superimposed on the system identification) that are introduced are mentioned.
\begin{itemize}
    \item The choice of \textit{model class structure} is essential for reducing over- and underfitting.
    The broad model class presented in Eqs.~\eqref{eq:model-G} and \eqref{eq:model-M} is used. 
    For the MSW incineration system identification, the domain of the time constants was limited to 
    \begin{equation}
        \forall i:\, T_{\str{p}i}\leq T_\str{p,max} = \qty{e4}{s} \punk{,} \label{eq:Tpmax}
    \end{equation}
    which was chosen as a time constant significantly larger than any of the dynamic effects that can be expected based on knowledge of the system.
    For each input, the number of poles in Eq.~\eqref{eq:model-G} is a hyperparameter.
    \item \textit{Additional inputs} allow additional parts of the system dynamics to be identified.
    By selectively extending the model class $\mathcal{M}$, this reduces underfitting.
    However, the inclusion of irrelevant inputs can increase overfitting.
    For each signal, the binary variable that decides whether to use the signal as model input is a hyperparameter.
    \item \textit{Sequential system identification}, used, e.g., by \citet{loosDynamicModelingCombustion1996}, aims to correctly attribute system dynamics to input signals by performing multiple consecutive stages of system identification. 
    Instead of identifying all models $G$ in Eq.~\eqref{eq:model-M} at once, the inputs are ranked and split to be used sequentially.
    In the first stage, only a selection of inputs is used. In the second stage, the already-identified parts remain unchanged and additional inputs are added, etc. 
    This reduces the number of degrees of freedom at any time in the identification algorithm and has proven effective in reducing overfitting \citep{loosDynamicModelingCombustion1996}.
    In this work, three sequential stages are used.
    For each input, the stage number in which the input is considered for the first time is a hyperparameter. 
    If it is known that a certain input is responsible for a major part of the system dynamics, it is always part of the first stage.
    \item \textit{L$_2$-Regularization} reduces overfitting by adding a penalty term of the form $\lambda\,\lVert \theta \rVert^2$ to Eq.~\eqref{eq:thetaopt}. This penalizes large $\theta$ values, reducing the complexity of the model \citep{ljungPerspectivesSystemIdentification2010}.
    For each stage, $\lambda$ is a hyperparameter.
\end{itemize}

A low $\MSE$ can be obtained if the sweet spot between overfitting and underfitting is found.
This requires good hyperparameter tuning, which can be defined as the tuning of hyperparameters $\eta$ influencing the model class structure or system identification optimization aiming to find 
\begin{equation} \label{eq:etaopt}
    \hat{\eta} = \argmin_{\eta \,\in\, D_\eta} J(\eta) \punk{.}
\end{equation}
In this equation, $D_\eta$ is the domain of $\eta$.
The procedure used to tune the hyperparameters is illustrated in \autoref{fig:overview-sysIDHP} and described below.

A prerequisite for hyperparameter tuning in this context is a performance metric which reflects the model performance on unseen data.
$K$-fold cross-validation is a popular method through which such metrics can be obtained \citep{arlotSurveyCrossvalidationProcedures2010}.
Given a set $\mathcal{E}$ of $N_E$ experiments which are available for system identification, divide the experiments in $K$ groups or folds, each group $k$ containing experiments $\mathcal{E}_k$.
For each fold $k \in \mathcal{K} = \{1,...,K\}$, find the optimal model $\mathcal{M}({\hat{\theta}_k^{\,\eta}})$ with
\begin{equation} \label{eq:thetaopt-Kfold}
    \hat{\theta}_k^{\,\eta} = \argmin_{\theta \,\in\, D_\theta} \sum_{i \,\in\, \mathcal{E} \setminus \mathcal{E}_k} \MSE(\theta|\eta)\big|_i \punk{,}
\end{equation}
in which $\MSE(\theta|\eta)\big|_i$ is the evaluation of $\MSE$ for experiment $i$ using parameters $\theta$ and hyperparameters $\eta$.
The evaluation metric $J(\eta)$ is defined as
\begin{equation}\label{eq:J-Kfold}
    J(\eta) = \frac{1}{N_E} \sum_{k \,\in\, \mathcal{K}} \sum_{i \,\in\, \mathcal{E}_k} \MSE({\hat{\theta}_k^{\,\eta}}|\eta) \big|_i \punk{.}
\end{equation}
Note that using $K$-fold cross-validation means that the optimization in Eq.~\eqref{eq:thetaopt-Kfold} needs to be done $K$ times, significantly increasing the evaluation time of the algorithm.

There are several methods for tuning hyperparameters, the most naive being grid search or random search, which test several sets of hyperparameters in a structured or random way.
If the computational cost of the underlying optimization is small and the hyperparameter space $D_\eta$ is low-dimensional with limited domain, these approaches are suitable.
The problem we consider, a multi-input system identification procedure of a complex system, has an expensive optimization function and $D_\eta$ is high-dimensional, with a large domain in some dimensions.
This makes naive search algorithms inefficient.
A more intelligent approach would use the already observed locations in $D_\eta$ and the observations made, which can be summarized as $\mathcal{D} = \left\{\left(\eta, J\left(\eta\right)\right)\right\}$, to determine the next $\eta$ to evaluate.

Bayesian Optimization (BO) is a method that uses past observations to select the new observation point.
It approaches the hyperparameter tuning task \eqref{eq:etaopt} as an optimization problem.
BO has been successfully applied in related fields, such as controller tuning \citep{brunzemaControllerTuningTimeVarying2022}.
In BO, the objective function $J(\eta)$ is modelled as a Gaussian process (GP) $\tilde{J}\left(\eta\,|\mathcal{D}\right)$ with $\mathcal{D}$ the set of past observation points and observed $J(\eta)$ \citep{garnettBayesianOptimization2023}.
Based on the GP, an acquisition function $\alpha$ is constructed, whose maximum dictates the next location in $D_\eta$ to observe.
The acquisition function contains a trade-off between exploiting regions of $D_\eta$ in which the expected value of $\tilde{J}\left(\eta\,|\mathcal{D}\right)$ is low, so that $J(\eta)$ is probably good, and exploring regions in which $\tilde{J}\left(\eta\,|\mathcal{D}\right)$ has high uncertainty, so that $J(\eta)$ is possibly good.
The BO algorithm is summarized in \autoref{alg:bo}.

\begin{algorithm}
\small
\caption{Bayesian Optimization algorithm \citep{garnettBayesianOptimization2023}.}
\label{alg:bo}
\begin{algorithmic}[1]

  \Require Initial dataset $\mathcal{D}$  \Comment{can be empty}
  \State $\hat{J_\eta}\leftarrow\infty$
  \Repeat
    \State \!\!\! Update GP $\tilde{J}$ and acquisition function $\alpha$
    \State \!\!\! $\eta \leftarrow \argmax_{\eta' \in D_\eta} \alpha\left(\eta'\,|\mathcal{D}\right)$ \Comment{get next location}
    \State \!\!\!$J_\eta \leftarrow J(\eta)$ \Comment{observe location (Eq.~\eqref{eq:J-Kfold})}
    \State \!\!\!$\mathcal{D} \leftarrow \mathcal{D} \cup \left\{\left(\eta, J_\eta\right)\right\}$ \Comment{update dataset}
    \State \!\!\!\textbf{if} $J_\eta < \hat{J}_\eta$: $\left(\hat{\eta}, \hat{J}_\eta\right) \leftarrow \left(\eta, J_\eta\right)$ \Comment{update best location}
  \Until{termination condition} \Comment{e.g. max iterations}
  \State \textbf{return} $\hat{\eta}$
\end{algorithmic}
\end{algorithm}

Although BO introduces an additional computational cost for the construction and evaluation of the Gaussian process and acquisition function, this cost is low compared to the cost to evaluate $J$, so that the targeted search for $\hat{\eta}$ is computationally less expensive than a naive search algorithm. 
By choosing a maximum number of 300 iterations as a termination condition, we ensure that the process could be completed in under 10 hours on a standard personal computer.
The BO hyperparameter tuning is implemented using built-in MATLAB functions, with the default ARD Matern 5/2 kernel for the Gaussian process and the `Expected Improvement' acquisition function $\alpha\left(\eta\,|\mathcal{D}\right)$, which evaluates how much the optimization objective $J$ is likely to improve if the next evaluation is done at $\eta$ given $\mathcal{D}$ \citep{themathworksinc.StatisticsMachineLearning2023}.
The MATLAB implementation allows for evaluating additional constraint functions, for example, to check the feasibility of a solution.
As a feasibility constraint, Eq.~\eqref{eq:Tpmax} was implemented; if the optimal model at the observation location has a pole with value $T_\str{p,max}$, the observation location is marked as infeasible.


\section{Model}\label{sec:model}
Two models of the MSW grate incineration plant are created. 
The first one, called Basic Model, applies the described system identification procedure with hyperparameter tuning (\sysIDHP) a single time on the complete closed-loop system. 
The second model divides the system into multiple connected subsystems that represent parts of the process.
Most of the subsystems are identified using the \sysIDHP{} method, some make use of algebraic equations instead.
The resulting model is called Comprehensive Model, as it offers more insight in the process by providing additional intermediate physical quantities as extra outputs.
Because the goal of the models is low-order modelling for unit control purposes, the models focus on the steam load dynamics.
The complete models are given in \autoref{sec:appFullModels}.

\subsection{Plant Data and Data Preprocessing}
Measurement data was available from a MSW plant in Europe with a rated steam capacity of \qty{32.2}{MW} at \qty{50}{bar} and \qty{430}{\degC}.
All signals discussed in the Introduction and given in \autoref{fig:grateInc-overview} are available for \num{10} consecutive days in \qty{5}{s} resolution.
The plant varies its operating point in a daily rhythm, between a steam load of \num{90} to \qty{100}{\%} during the day and a part load of circa \qty{65}{\%} during the night.
Since the setpoint is changed manually by the plant operator, the exact transition between the regimes varies from day to day, sometimes consisting of a few large setpoint changes, sometimes of several small changes spread over a large or a small time span.
Minor changes to the $\Ot$ setpoint (also operated manually) often occur together with a change in the setpoint of the steam load.
The dynamics associated with these $\Ot$ setpoint changes can not be observed separately.
Therefore, the $\Ot$ setpoint is not considered as input in this work.

The frequent changes in the load setpoint make the data particularly suitable for testing the proposed system identification method.
It should be noted that this operation method is not common for waste incineration plants, which, if sufficient waste is available and no regulatory restrictions apply, normally run continuously at full load.
However, any waste incineration plants could ramp up and down the steam load, providing data similar to the available data.
Also, the high-load parts of start-up and shut-down processes can provide similar data, if the auxiliary burners of the plant are not in operation.

Data standardization of the form
\begin{equation}\label{eq:standard}
    x^* = \frac{x-\bar{x}}{\max(x)-\min(x)}
\end{equation}
is used for all measurement signals. 
Not only does this compensate offsets, it also maps all signals to a similar range, avoiding signals with a larger magnitude from dominating the \MSE.

The dataset needs to be split into multiple experiments, which are assigned to two groups, the training data and the test data, in agreement with \autoref{fig:overview-sysIDHP}.
The data is split and assigned as follows:
\begin{enumerate}
    \item The data is split automatically based on the steam load setpoint, splitting $10$ minutes before a major setpoint change, resulting in $27$ experiments.
    \item The experiments are randomly split in $20$ training and $7$ validation experiments.
    \item The training experiments are randomly split into $K=5$ groups for cross-validation, so that each group consists of $4$ experiments.
\end{enumerate}

The measurement of the position of the ram feeder, $x_\str{ram}$, is transformed to a variable proportional to the volumetric fuel flow, $\VWaste$.
The position of the ram varies periodically between far from and close to the grate. 
We split this movement in periods which start when the ram feeder is at its farthest point from the grate at time $t_i$.
$t_{i+1}$ is the next time such a minimum position is observed.
The volumetric fuel flow can then be approximated from $x_\str{ram}$ and the surface area of the ram, $A_\str{ram}$, as
\begin{multline} \label{eq:stoker}
    \VWaste(t) = A_\str{ram} \frac{\max_{\tau \in [t_i,t_{i+1}]}(x_\str{ram}(\tau)) - x_\str{ram}(t_i)}{t_{i+1}-t_i} \\ \forall t \in [t_i,t_{i+1}] \punk{.}
\end{multline}
Because of the data standardization (Eq.~\eqref{eq:standard}), it is not necessary to know $A_\str{ram}$ to be able to use $\VWaste$.
Due to the varying density of the waste, the identified correlation cannot be extended to the fuel mass flow. 
The reference plant has two parallel ram feeders, so that $\VWaste$ is calculated for each ram feeder individually, and the average $\VWaste$ is used.

The steam load of the plant can be expressed as the mass flow rate, $\mLS$, or the thermal power, $\Qsteam$.
The two are related through
\begin{equation}
    \Qsteam = \mLS \left(h_\str{s} - h_\str{feedwater}\right) \punk{.}
\end{equation}
The thermal power formulation is used in this work, because it is more robust, as the nonlinear dependency of the water enthalpy on pressure and temperature is compensated.
An additional advantage of using $\Qsteam$ instead of $\mLS$ is the generalizability of the approach, as $\Qsteam$ will be less sensitive to changes in operating mode (for example, between plants which have sliding pressure and constant pressure mode).

\subsection{Basic Model}
\subsubsection{Setup}
The basis of the system to be identified is
\begin{equation}
    \Qsteam = G \left(\QsteamSP \right)\punk{.}
\end{equation}
From our knowledge of the process, variables that might influence the system are made available as system inputs, which the \sysIDHP{} might choose to use.
The fuel composition is known to be a disturbance with a strong effect on the steam load.
Although not measured, information about the composition of the fuel on the grate is contained in the volumetric concentration measurements of the flue gas at the end of the boiler.
The measured concentrations of \ce{CO2}, \ce{H2O}, and \ce{O2} are linked to the combustion processes shortly before, and the \ce{H2O} measurement is also related to the fuel drying process.
If steady-state conditions were present, these measurements, in combination with air flow measurements, could be used to determine the fuel composition \citep{davidDeterminationFuelComposition2023}.
Dynamically, this is not possible.
However, $\COt$, $\HtO$ and $\Ot$ still contain additional information about the current combustion process, under the assumption that the transport time of the flue gas between the grate and the boiler end is small compared to the time span in which the fuel composition on the grate undergoes significant changes.
This assumption is justified, considering that the residence time of the solids on the grate is typically between half an hour and one hour \citep{buekensIncinerationTechnologies2013}.
It is also known that the temperature of the preheated primary air $\TPair$ has an effect on the fuel drying process and the steam load.

The resulting system to be identified is given as
\begin{subequations}
\begin{align}
    \Qsteam &= G\left(\QsteamSP, \,\TPair, \,\HtO, \,\COt, \,\Ot\right) \\
    &= G\left(\QsteamSP, \,\delta\right) \punk{,}
\end{align}
\end{subequations}
in which $\delta$ summarizes the additional inputs.

\subsubsection{Results}
\begin{figure}[htbp]
    \centering
    \begingroup\endlinechar=-1
    \resizebox{0.6\textwidth}{!}{\input{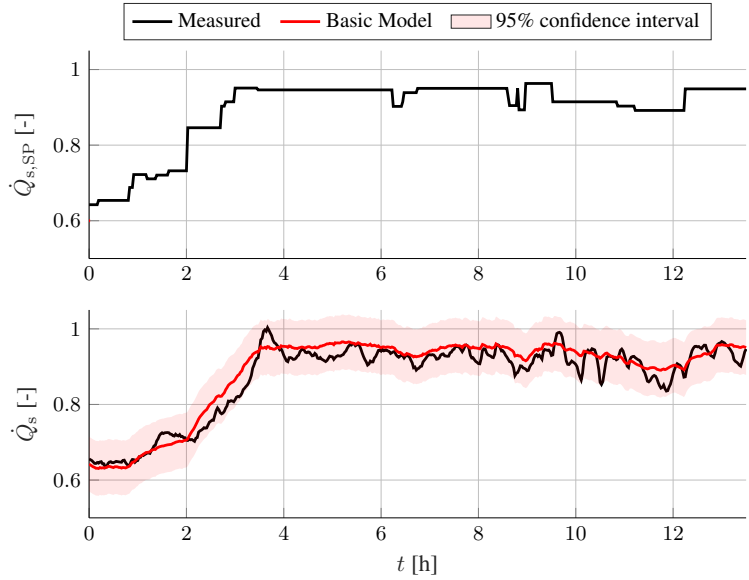}}
    \endgroup
    \caption{Validation experiment $1/7$. Top: dimensionless steam load setpoint. Bottom: measured steam load and Basic Model output with \qty{95}{\%} confidence interval\protect\footnotemark.}
    \label{fig:basic-res}
\end{figure}

The \sysIDHP{} procedure was run and found the best performance using all 5 inputs in a single stage, with one pole for inputs $\QsteamSP$, $\TPair$ and $\Ot$ and two poles for $\HtO$ and $\COt$.
The resulting $\Rt$ was \qty{91.9}{\%} on the training and \qty{91.5}{\%} on the test data, showing that the presented procedure is well suited for this complex system.
The resulting output for one of the seven test/validation experiments is shown in \autoref{fig:basic-res}. 
It is seen that the general behaviour of the complex system is well-modelled.
Based on the covariance of the estimated system parameters, the \qty{95}{\%} confidence interval of the identified system is also shown, which always encompasses the measured signal.

\footnotetext{Dimensionless means relative to a reference value; a dimensionless value of 1 corresponds to the reference value. See \autoref{tab:offsetsetc} for the used references.}

Depending on the application, one might be interested in using the model as a multistep ahead predictor, in which case the additional model inputs $\delta$ would not be available for the time within the prediction horizon.
In this case, a good estimate is to use a rolling mean of previous $\delta$ measurements as input.
One could also leave out the influence of the additional inputs $\delta$ completely, by setting the standardized input to 0.

\subsection{Comprehensive Model}
\subsubsection{Setup}

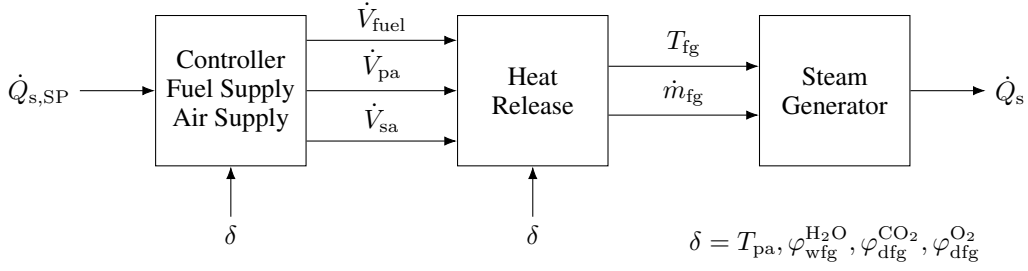
\begin{figure*}[tbhp]
    \centering
        \begin{tikzpicture}[
        node distance = 6.7mm and 10mm,
        block/.style={draw, minimum size=20mm, align=center},
        thinblock/.style={draw, minimum size=20mm, minimum height = 13mm, align=center}
    ]
    \node (process1) [block] {Controller\\Fuel Supply\\Air Supply};
    \coordinate[left=of process1.west, label=left:$\QsteamSP$] (QsSP);
    
    \coordinate[below=of process1.south, label=below:$\delta$] (delta1);
    \coordinate[above right=of process1.east, label=above:$\VWaste$] (VWaste);
    \coordinate[right=of process1.east, label=above:$\VPair$] (VPair);
    \coordinate[below right=of process1.east, label=above:$\VSair$] (VSair);

    \node (process2) [block, right=of VPair] {Heat\\Release};
    \coordinate[below=of process2.south, label=below:$\delta$] (delta2);
    \coordinate[right=of process2.east, label=above:$\Tfurn$, yshift = 3.25mm] (Tfurn);
    \coordinate[right=of process2.east, label=above:$\mfurn$, yshift = -3.25mm] (mfurn);

    \node (process3) [block, right=of Tfurn.south, yshift = -3.25mm] {Steam\\Generator};
    \coordinate[right=of process3.east, label=right:$\Qsteam$] (Qs);

    \node (test) [ below=of process3.south] {$\delta = \TPair, \HtO, \COt, \Ot$};

    \draw[-Latex] (QsSP) -- (process1.west |- QsSP);
    \draw[-Latex] (delta1) -- (process1.south);
    
    \draw[-Latex] (delta2) -- (process2.south);
    \draw[-Latex] (VWaste -| process1.east) -- (VWaste -| process2.west);
    \draw[-Latex] (VPair -| process1.east) -- (VPair -| process2.west);
    \draw[-Latex] (VSair -| process1.east) -- (VSair -| process2.west);
    
    \draw[-Latex] (Tfurn -| process2.east) -- (Tfurn -| process3.west);
    \draw[-Latex] (mfurn -| process2.east) -- (mfurn -| process3.west);
    
    \draw[-Latex] (Qs -| process3.east) -- (Qs);

    \end{tikzpicture}
    \caption{Schematic structure of subprocesses and measurement signals in a waste incineration plant.}
    \label{fig:compreh-schem}
\end{figure*}

A disadvantage of the Basic Model is the lack of insight in the process, notably the lack of information about the flue gas parameters $\Tfurn$ and $\mfurn$. 
It is possible to extend the Basic Model to include these and other variables of interest as additional outputs.
However, with the used process model structure (Eqs.~(\ref{eq:model-G}),(\ref{eq:model-M})), there would not be any coupling between the outputs.
This might lead to unphysical results and would not make optimal use of the available measurement signals.
State-space models could overcome this problem, but introduce a large amount of extra degrees of freedom, increasing the risk of overfitting.
This was observed when attempting identification using state-space models.

As an alternative, it is possible to use measurements taken at different locations in the process to identify different subprocesses, after which these can be coupled in series.
Dynamically, the behaviour of the closed-loop plant can be split into the incineration controller, the fuel and air supply, the heat release, and the steam generation.
These subprocesses and their inputs and outputs are illustrated schematically in \autoref{fig:compreh-schem}.
Due to the availability of signals in the closed-loop identification setup, the controller cannot be identified separately, but can be identified together with the fuel and air supply.

The models for each output of the subprocesses can be found using the \sysIDHP{} method.
Knowledge of the physical system can sometimes be used instead, to algebraically formulate the output quantity as a function of the input or to construct additional meaningful inputs from available signals.

The volumetric fuel flow, which is an output of the first subprocess, would be proportional to the steam load setpoint if fuel properties were constant. 
It can be modelled as
\begin{equation} \label{eq:VWaste-prop}
    \VWaste = K_\str{P}\,\QsteamSP + \dVWaste \punk{,}
\end{equation}
in which $K_\str{P}$ is a proportionality constant that can be identified and $\dVWaste$ contains deviations from the proportional relationship, which originate from the unknown variations of fuel properties and variations in the stoker efficiency (which could be expressed as variations in effective $A_\str{ram}$ in Eq.~\eqref{eq:stoker}).

The flue gas mass flow rate $\mfurn$ can also be expressed algebraically in function of measurement signals.
This is done by balancing the volumetric flow rates of the air with the flue gas concentration measurements, similar to the balancing done by \citet{beckmannPossibilitiesProcessOptimization2005}.
This requires the assumption that the dynamics of air and flue gas travelling through, respectively, the grate and the boiler, are fast compared with the main system dynamics (the incineration process and heat transfer in the steam generator).
The air at the inlet is taken as a mixture of \ce{O2} and \ce{N2}, which is assumed to be inert, and it is assumed that the only substantial constituents of the flue gas are \ce{N2}, \ce{O2}, \ce{CO2}, and \ce{H2O}, so that
\begin{equation} \label{eq:N2wfg}
    \varphi^{\ce{N2}}_\str{wfg} = 1-\varphi^{\ce{H2O}}_\str{wfg} -\varphi^{\ce{O2}}_\str{wfg} -\varphi^{\ce{CO2}}_\str{wfg} \punk{,}
\end{equation}
in which the wet flue gas concentration of \ce{O2} is calculated as
\begin{equation}
    \varphi^{\ce{O2}}_\str{wfg} = \varphi^{\ce{O2}}_\str{dfg}
    \left(1-\varphi^{\ce{H2O}}_\str{wfg}\right)
\end{equation}
and analogously for \ce{CO2}.
The wet flue gas volumetric flow rate under standard conditions, $\dot{V}_\str{fg}$ and the flue gas mass flow are given by
\begin{equation}
    \dot{V}_\str{fg} = \frac{\varphi^{\ce{N2}}_\str{air}}{\varphi^{\ce{N2}}_\str{wfg}} \left( \VPair + \VSair \right) \punk{,} \label{eq:Vwfg}
\end{equation}
\begin{equation}
    \mfurn = \dot{V}_\str{fg} \, V_\str{m} \, \sum_{\substack{i \in  \{\ce{N2}, \ce{O2},\\ \ce{H2O}, \ce{CO2}\} }} \varphi_\str{wfg}^i \, \text{MM}_i \punk{.} \label{eq:mfg}
\end{equation} 
    
Here, $\text{MM}_i$ is the molar mass of component $i$ and $V_\str{m}$ is the molar volume of an ideal gas under standard conditions.
This algebraic method does not need any parametrization and can be used without having a measurement of $\mfurn$ available.

Additional inputs are made available to the \sysIDHP{} procedure, as was done for the Basic Model.
Because the volumetric flue gas concentrations $\COt$, $\HtO$, and $\Ot$ contain information about the fuel properties in the combustion process, they are taken as inputs for the `heat release'-subprocess. 
They are also taken as inputs for the first subprocess, where the controller might have access to these measurements and could act on them.
The primary air temperature, $\TPair$, which can influence the combustion processes, is also taken as an additional input for these subprocesses.
For the `steam generation'-subprocess, a different additional input, namely,
\begin{equation} \label{eq:fgprod}
    \Gamma = \Tfurn \, \mfurn \punk{,}
\end{equation}
is available.
This is motivated based on the relation
\begin{subequations}\label{eq:Qhm}
\begin{align}
    \Qsteam &= \Delta h_\str{s} \, \mLS \\
            &= \Delta h_\str{wfg} \, \mfurn \\
            &= c_\str{p} \, \left(T_\str{wfg,in}-T_\str{wfg,out}\right) \, \mfurn \punk{,}
\end{align}
\end{subequations}
in which $c_\str{p}$ is the average specific heat capacity of the flue gas and indices `in' and `out' stand for the beginning and ending of the flue gas passes in the steam generator.
In Eq.~\eqref{eq:Qhm}, the product of the flue gas temperature and mass flow rate is proportional to $\Qsteam$, motivating the use of $\Gamma$.

\subsubsection{Results}

The \sysIDHP{} procedure was performed for all outputs of the different subprocesses shown in \autoref{fig:compreh-schem}, except for the volumetric fuel flow, for which Eq.~\eqref{eq:VWaste-prop} was parametrized, and the flue gas mass flow, for which Eq.~\eqref{eq:mfg} was used.
For each of the identified models, the \sysIDHP{} found hyperparameter combinations that perform significantly better than the combinations that were provided to the BO algorithm (\autoref{alg:bo}) as the initial data set $\mathcal{D}$. 
The initial dataset consisted of educated guesses, ranging from very simple models with few of the available inputs used to elaborate models, such as second-order systems, for all available inputs. 
Some optimal models did not include all available inputs or used multiple stages.
This shows that the procedure correctly selects and ranks inputs.
If desired, the number of used inputs can be further reduced by reducing the maximum allowed time constant in Eq.~\eqref{eq:Tpmax}.

Model performance metrics for the Subprocess Models, as well as for the Basic and complete Comprehensive Model (the complete configuration shown in \autoref{fig:compreh-schem}), are shown in \autoref{tab:metrics}.
It is seen that the Comprehensive Model is slightly inferior to the Basic Model with respect to overall performance.
However, the Subprocess identification works very well considering the complexity of the process and disturbance influences.
The overall high $\Rt$ and low $\MSE$ indicate that there is no underfitting.
The small differences in performance between the training and the test data indicate that there is no overfitting either.

\begin{table}[htbp]
    \centering
    \caption{Model performance metrics for Basic and Comprehensive (Comp.) Model, as well as for the Comp. Subprocess Models. The $\MSE$ is given for the standardized variables (Eq.\eqref{eq:standard}).\vspace{1ex}}
    \begin{tabular}{c|cc|cc}              & \multicolumn{2}{c|}{Estimation} & \multicolumn{2}{c}{Validation} \\              & $\MSE$ [-]             & $\Rt$ [\%]            & $\MSE$ [-]           & $\Rt$ [\%]           \\ \hline Basic         & 2.46\e{-3}              & 91.91            & 2.61\e{-3}             & 91.53            \\ \hline Comp. & 3.96\e{-3}              & 87.36            & 3.71\e{-3}             & 88.12            \\ \hline \VPair         & 3.69\e{-3}              & 72.33            & 3.76\e{-3}             & 70.87            \\ \VSair         & 3.84\e{-3}              & 76.59            & 4.99\e{-3}             & 71.87            \\ \Tfurn         & 3.81\e{-3}              & 76.94            & 4.27\e{-3}             & 81.52            \\ \Qsteam        & 2.72\e{-3}              & 90.91            & 3.40\e{-3}             & 89.30           \end{tabular}
    \label{tab:metrics}
\end{table}

The first and second subprocess identifications perform the worst and have the largest performance differences between estimation and validation.
This can be explained by the strong nonlinearities in these processes that contain the incineration processes and controller.
The steam generator subprocess is less complex.
This is reflected in better performance of the identified system.
It is remarkable that the performance of the complete comprehensive model is similarly good as the last subprocess.
This can be explained by the low-pass behaviour of the last subprocess.
Thanks to the inertia of the steam generator, fast fluctuations or inaccuracies of the previous subprocesses are smoothed out.

\begin{figure}[htbp]
    \centering
    \begingroup\endlinechar=-1
    \resizebox{0.6\textwidth}{!}{\input{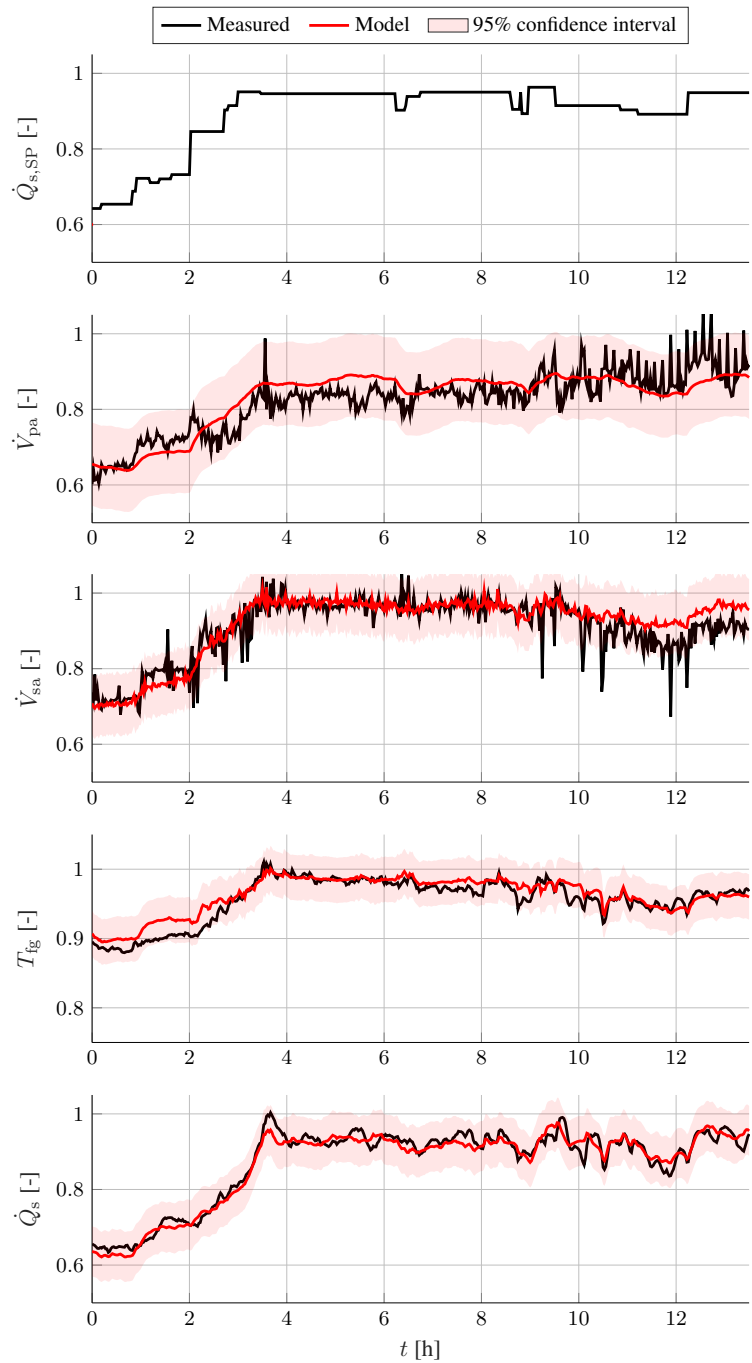}}
    \endgroup
    \caption{Validation results comparing the dimensionless Comprehensive Subprocess Model outputs and confidence intervals with corresponding measurements (validation experiment $1/7$).}
    \label{fig:comp-sub-res}
\end{figure}

The performance of the Comprehensive Model subprocesses is also illustrated for one validation experiment in \autoref{fig:comp-sub-res}.
It is clear that the models capture the main dynamics of each of the subprocesses.
The fast dynamics in the volumetric air flows are not captured by the models.
These dynamics might be associated with soot blowing and cannot be modelled with the model inputs used.
The $95\%$ confidence interval of these models does encompass most of the measured points.
The furnace temperature and steam load, shown in \autoref{fig:comp-sub-res}, have significantly smaller confidence intervals than the air flow models, also encompassing most of the measurements.
This shows that the identification procedure works excellently for the identified processes: both the model prediction and the confidence interval are meaningful for unseen validation (i.e.~test) data.
It can also be seen that some fast dynamics are captured, which is thanks to the additional inputs considered in the identification procedure.

\section{Model Comparison and Discussion}\label{sec:discussion}

All models are successful in predicting their respective outputs, as seen in \autoref{tab:metrics}, \autoref{fig:basic-res} and \autoref{fig:comp-sub-res}.
They also provide significant confidence intervals, conveying information about the uncertainty associated with the varying fuel composition and the process itself.

\begin{figure}[htbp]
    \centering
    \begingroup\endlinechar=-1
    \resizebox{0.6\textwidth}{!}{\input{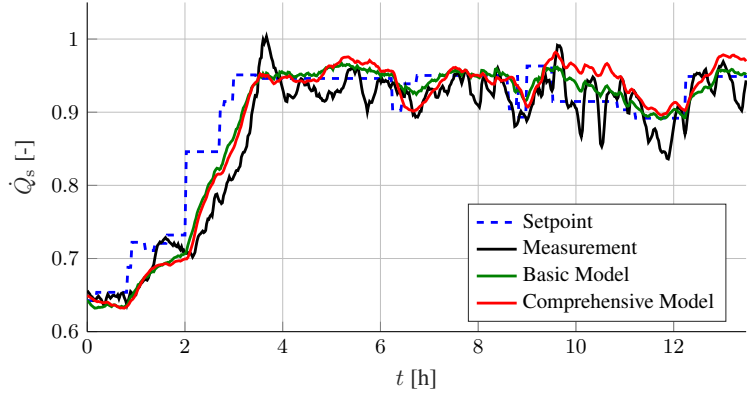}}
    \endgroup
    \caption{Validation results showing the dimensionless steam load setpoint, corresponding measurement and the output of both the Basic and the complete Comprehensive Model (validation experiment $1/7$).}
    \label{fig:both-res}
\end{figure}

Validation results of both the Basic and Comprehensive Model are shown in \autoref{fig:both-res}, together with the setpoint and measurement of the steam load.
Both models agree well with the measurement data and give very similar results, both at times with major setpoint changes and times when disturbances dominate the dynamic behaviour.
The similar performance indicates that the coupling of multiple subprocesses using process knowledge and the use of additional sensor measurements (as was done for the Comprehensive Model) does not deteriorate, nor improve the model quality.

\begin{figure}[htbp]
    \centering
    \begingroup\endlinechar=-1
    \resizebox{0.5\textwidth}{!}{\input{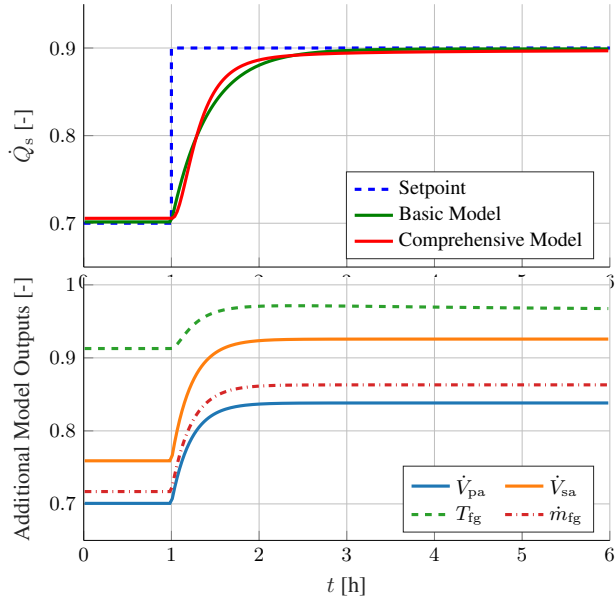}}
    \endgroup
    \caption{Simulated step on the steam load setpoint with constant additional inputs $\delta$. Top: Steam load setpoint and model responses. Bottom: additional outputs of the Comprehensive Model.}
    \label{fig:step-QSP}
\end{figure}

The results of a simulated step in the steam load setpoint are shown in \autoref{fig:step-QSP}.
It is seen that the Basic and Comprehensive model have very similar responses, with the characteristic slow response, typical for grate-firing.
The inlet air flows already show these slow dynamics, which means that the main inertia of the model is located in the first subprocess.
This is consistent with field knowledge and other research \citep{bauerModellingGrateCombustion2010};
because of the long fuel conversion times on the grate, the air flows slowly react on a changing steam load (and changing fuel supply).

The consistent performance and results of the Basic Model and the Comprehensive Model are positive, and indicate that the Basic Model should be used if only the steam load output is required.
The Comprehensive Model is more complex to create, but offers additional outputs which match the measured variables well for validation data.
Depending on the specific control application, the additional outputs can be useful, e.g. for monitoring the flue gas temperature while a model-based controller plans an operating schedule.

\section{Final Remarks}\label{sec:final}
The proposed system identification method with hyperparameter tuning successfully combines system identification with multiple methods that find their origin in machine learning: cross-validation, regularization and Bayesian optimization.
Using the method, it is possible to identify the closed-loop dynamics of grate-fired municipal solid waste (MSW) incineration plants, which are known for their heavy disturbances resulting from the fluctuating fuel composition.
This implies that the same method could be used for other complex processes or systems with heavy disturbances.
The combination of sequential system identification with Bayesian optimization could be particularly useful for the closed-loop identification of over-actuated systems.
The use of Bayesian optimization for hyperparameter tuning looks promising even for other system identification procedures, as in this work it systematically outperformed estimated guesses in the selection of inputs and model structure.

The combination of the system identification method with physics-based knowledge to yield additional outputs, as was done for the Comprehensive Model, did not improve, nor deteriorate, the model performance.
The main advantage of this method is thus the availability of additional model outputs, which must be weighed against the additional modelling effort.
However, making additional signals available as inputs for the models improved model performance, as these signals were chosen by the Bayesian optimization to be part of the final model.

The system identification in this work focused on the dynamics of the steam load in a closed-loop identification procedure.
The method can be extended to include, for example, the dynamics of oxygen setpoint changes or open-loop identification, if the appropriate data is available.

The presented method can be easily applied to industrial MSW grate incineration plants to obtain a low-order closed-loop system model, which cannot be achieved by any literature approach known to the authors.
The models can be used as part of unit control schemes, allowing for a more flexible operation of MSW incineration plants, adapted to the needs of the power system.

Another possible application is the evaluation of the current condition of existing incineration plants:
models could be created for different plants, and by using the model responses and confidence intervals, experts could evaluate whether a plant could profit from different operating methods, maintenance, or a retrofit, e.g., when observing exceptionally slow dynamics or large confidence intervals.
The developed models, which were created in order to evaluate the method, are disclosed in \autoref{sec:appFullModels} and can also be used for further research, the development of unit control schemes, the qualitative assessment of the influence of fuel variations on process outputs, or the combination with steam cycle models such as \citep{lipsDynamicModellingSimulation2022}.

\section{CReDiT Author Statement}
\textbf{Johannes Lips}: Conceptualization, Methodology, Validation, Writing -- Original Draft.
\textbf{Stefan \mbox{DeYoung}}: Data Curation, Writing -- Review and Editing.
\textbf{Max Schönsteiner}: Data Curation, Writing -- Review and Editing.
\textbf{Hendrik Lens}: Supervision, Writing -- Review and Editing.

\section{Acknowledgements}
This research is part of the project ``CHP 4.0 -- Regional combined heat and power plants in a changing energy system.'' It is funded by the German Federal Ministry for Economic Affairs and Climate Action under grant number FKZ 03EE5031 E.

\appendix
\setcounter{table}{0}
\section{Identified Models} \label{sec:appFullModels}
All models are identified using standardized variables (Eq.~\eqref{eq:standard}), which require scaling and offsetting.
For easier use in independent research, the full models are given below using offset dimensionless variables as in- and outputs.
As before, dimensionless variables are defined as normalized on a reference value, so that their typical regular range is $[0,1]$.
For concentrations, \qty{100}{\%} is taken as a reference.
As concentrations of \ce{CO2}, \ce{O2}, and \ce{H2O} are normally one order of magnitude smaller than \qty{100}{\%}, the gains associated with these inputs are sometimes high without this indicating a major influence of this input on the output.
A list of variables, offsets and reference values as used in this work is given in \autoref{tab:offsetsetc}.
The in-/ output $x^\bullet$ of the system relates to the dimensionless variable $x$ as
\begin{equation}
    x^\bullet = x - \Delta_c \punk{.}
\end{equation}
The transformation between $x^\bullet$ and $x$ needs to be done before and after applying the model.
Inputs that were available to the \sysIDHP{} procedure, but not selected as part of the final model, are not included in the model description given below.
Although dead times $T_\str{d}$ were possible in the model structure \eqref{eq:model-G}, any identified delays were negligible compared with the governing system dynamics, and they are left out in the model descriptions, i.e.~$T_\str{d}=0$, resulting in fully linear models.

\begin{table}[tbph]
\centering
\caption{Model variables, offsets, reference values and units.}
\begin{tabular}{c|c|rl}
Variable           & Offset $\Delta_c$    & Reference & Units  \\ \hline
$\Qsteam$            & 0.817581  & 32.2        & MW     \\
$\VPair$             & 0.781564  & \num{3.95e4}  & \Nmc   \\
$\VSair$             & 0.857033  & \num{1.76e4}  & \Nmc   \\
$\TPair$             & 1.029000  & 120       & \degC  \\
$\Tfurn$             & 0.944238  & 880       & \degC  \\
$\mfurn$             & 0.802818  & 23        & kg/s   \\
$\Gamma$             & 0.758052  & $2.02\e{4}$  & K kg/s\\
$\Ot$                & 0.078830  & 100       & \%     \\
$\COt$               & 0.112913  & 100       & \%     \\
$\HtO$               & 0.144002  & 100       & \%     \\
\end{tabular}
\label{tab:offsetsetc}
\end{table}

\subsection{Basic Model}
The Basic Model takes the form
\begin{equation} \label{eq:transfer}
    Y = G U \punk{,}
\end{equation}
with 
\begin{align*}
    y = &\,\Qsteam \punk{,}\\
    u = &\begin{bmatrix} \QsteamSP & \TPair & \HtO & \COt & \Ot \end{bmatrix}^\str{T} \punk{,} \\
    G(s) = &\left[\begin{matrix} 
    \frac{0.98703}{1525.2s+1} &
    \frac{0.082006}{38.561s+1} &
    \frac{0.10116}{\left(566.98s+1\right)^2}
    \end{matrix}\right.\\
    &\left.\begin{matrix} 
    \frac{-0.7977}{\left(7635.1s+1\right)^2} &
    \frac{-1.4063}{320.17s+1}
    \end{matrix}\right] \punk{.}
\end{align*}

\subsection{Comprehensive Model}
The subprocesses of the Comprehensive Model take the same form as the Basic Model.
The primary air volumetric flow model is given by:
\begin{align*}
    y = &\,\VPair \punk{,} \\
    u = &\begin{bmatrix}\QsteamSP& \TPair& \HtO & \COt & \Ot \end{bmatrix}^\str{T} \punk{,} \\
    G(s) = &\left[\begin{matrix}
    \frac{0.68754}{799.67s+1} &
    \frac{0.33679}{4025.5s+1} &
    \frac{0.89669}{836.38s+1}
    \end{matrix}\right.\\
    &\left.\begin{matrix}
    \frac{-0.75509}{\left(4686.5s+1\right)^2} &
    \frac{-0.79057}{5182s+1}
    \end{matrix}\right] \punk{.}
\end{align*}

The secondary air volumetric flow model is given by:
\begin{align*}
    y = &\,\VSair \punk{,} \\
    u = &\begin{bmatrix}\QsteamSP & \HtO & \COt & \Ot \end{bmatrix}^\str{T} \punk{,} \\
    G(s) = &\begin{bmatrix}
    \frac{0.83353}{830.16s+1} &
    \frac{-0.17113}{96.153s+1} &
    \frac{-1.4383}{\left(8367.7s+1\right)^2} &
    -1.271
    \end{bmatrix} \punk{.}
\end{align*}

The flue gas temperature model is given by:
\begin{align*}
    y = &\,\Tfurn \punk{,} \\
    u = &\begin{bmatrix}\VPair & \VSair &  \QsteamSP & \HtO
    \end{bmatrix}^\str{T} \punk{,} \\
    G(s) = &\left[\begin{matrix} 
    \frac{-0.057537}{\left(6113s+1\right)\left(4262.9s+1\right)} &
    \frac{0.34828}{\left(213.55s+1\right)\left(3.638s+1\right)}
    \end{matrix}\right.\\
    &\left.\begin{matrix} 
    \frac{0.016756}{\left(2114.7s+1\right)\left(2004.7s+1\right)} &
    0.0041691
    \end{matrix}\right] \punk{.}
\end{align*}

The flue gas mass flow model is analytically described in Eq.~\eqref{eq:mfg}.

The product of flue gas temperature and mass flow, $\Gamma$, is defined in Eq.~\eqref{eq:fgprod} and used as input for the steam generator model.
The identified steam  generator model is given by:
\begin{align*}
    y = &\,\Qsteam \punk{,} \\
    u = &\begin{bmatrix}\Gamma & \mfurn &  \Tfurn \end{bmatrix}^\str{T} \punk{,} \\
    G(s) = &\begin{bmatrix} 
    \frac{0.81495}{(189.63\,s+1)\,(181.22\,s+1)} &
    \frac{0.14197}{9479.9\,s+1} &
    \frac{0.67983}{30.29\,s+1}
    \end{bmatrix} \punk{.}
\end{align*}






\bibliographystyle{unsrtnat} 
\bibliography{BayesOptArticle.bib}

\begin{thebibliography}{25}
\providecommand{\natexlab}[1]{#1}
\providecommand{\url}[1]{\texttt{#1}}
\expandafter\ifx\csname urlstyle\endcsname\relax
  \providecommand{\doi}[1]{doi: #1}\else
  \providecommand{\doi}{doi: \begingroup \urlstyle{rm}\Url}\fi

\bibitem[Makarichi et~al.(2018)Makarichi, Jutidamrongphan, and Techato]{makarichiEvolutionWastetoenergyIncineration2018}
Luke Makarichi, Warangkana Jutidamrongphan, and Kua-Anan Techato.
\newblock The evolution of waste-to-energy incineration: {{A}} review.
\newblock \emph{Renewable and Sustainable Energy Reviews}, 91:\penalty0 812--821, August 2018.
\newblock ISSN 1364-0321.
\newblock \doi{10.1016/j.rser.2018.04.088}.

\bibitem[Qin et~al.(2008)Qin, Bai, Shen, and Zhang]{qinDesignCombustionControl2008}
Yufei Qin, Yan Bai, Zhongli Shen, and Keming Zhang.
\newblock Design of {{Combustion Control System}} for {{MSW Incineration Plant}}.
\newblock In \emph{2008 {{International Conference}} on {{Intelligent Computation Technology}} and {{Automation}} ({{ICICTA}})}, pages 341--344, {changsha, Hunan}, October 2008. {IEEE}.
\newblock ISBN 978-0-7695-3357-5.
\newblock \doi{10.1109/ICICTA.2008.309}.

\bibitem[Leskens et~al.(2002)Leskens, Van~Kessel, and {Van den Hof}]{leskensMIMOClosedloopIdentification2002}
M.~Leskens, L.B.M. Van~Kessel, and P.M.J. {Van den Hof}.
\newblock {{MIMO}} closed-loop identification of an {{MSW}} incinerator.
\newblock \emph{Control Engineering Practice}, 10\penalty0 (3):\penalty0 315--326, March 2002.
\newblock ISSN 09670661.
\newblock \doi{10.1016/S0967-0661(01)00139-3}.

\bibitem[Magnanelli et~al.(2020)Magnanelli, Tran{\aa}s, Carlsson, Mosby, and Becidan]{magnanelliDynamicModelingMunicipal2020}
Elisa Magnanelli, Olaf~Lehn Tran{\aa}s, Per Carlsson, Jostein Mosby, and Michael Becidan.
\newblock Dynamic modeling of municipal solid waste incineration.
\newblock \emph{Energy}, 209:\penalty0 118426, October 2020.
\newblock ISSN 03605442.
\newblock \doi{10.1016/j.energy.2020.118426}.

\bibitem[Alobaid et~al.(2018)Alobaid, {Al-Maliki}, Lanz, Haaf, Brachth{\"a}user, Epple, and Zorbach]{alobaidDynamicSimulationMunicipal2018}
Falah Alobaid, Wisam Abed~Kattea {Al-Maliki}, Thomas Lanz, Martin Haaf, Andreas Brachth{\"a}user, Bernd Epple, and Ingo Zorbach.
\newblock Dynamic simulation of a municipal solid waste incinerator.
\newblock \emph{Energy}, 149:\penalty0 230--249, April 2018.
\newblock ISSN 0360-5442.
\newblock \doi{10.1016/j.energy.2018.01.170}.

\bibitem[Xia et~al.(2020)Xia, Shan, Chen, Du, Huang, and Bai]{xiaTwofluidModelSimulation2020}
Zihong Xia, Peng Shan, Caixia Chen, Hailiang Du, Jie Huang, and Li~Bai.
\newblock A two-fluid model simulation of an industrial moving grate waste incinerator.
\newblock \emph{Waste Management}, 104:\penalty0 183--191, March 2020.
\newblock ISSN 0956-053X.
\newblock \doi{10.1016/j.wasman.2020.01.016}.

\bibitem[{VDI/VDE-Gesellschaft Mess- und Automatisierungstechnik (GMA)}(2003)]{vdi-gesellschaftVDIVDE35082003}
{VDI/VDE-Gesellschaft Mess- und Automatisierungstechnik (GMA)}.
\newblock {VDI/VDE 3508 Unit control of thermal power stations}.
\newblock {VDI/VDE-Richtlinie}, {VDI-Verlag}, {D{\"u}sseldorf}, September 2003.

\bibitem[Bauer et~al.(2010)Bauer, Gölles, Brunner, Dourdoumas, and Obernberger]{bauerModellingGrateCombustion2010}
Robert Bauer, Markus Gölles, Thomas Brunner, Nicolaos Dourdoumas, and Ingwald Obernberger.
\newblock Modelling of grate combustion in a medium scale biomass furnace for control purposes.
\newblock \emph{Biomass and Bioenergy}, 34\penalty0 (4):\penalty0 417--427, April 2010.
\newblock ISSN 09619534.
\newblock \doi{10.1016/j.biombioe.2009.12.005}.

\bibitem[Kortela and {Jamsa-Jounela}(2015)]{kortelaFaulttolerantModelPredictive2015}
Jukka Kortela and Sirkka-Liisa {Jamsa-Jounela}.
\newblock Fault-tolerant model predictive control ({{FTMPC}}) for the {{BioGrate}} boiler.
\newblock In \emph{2015 {{IEEE}} 20th {{Conference}} on {{Emerging Technologies}} \& {{Factory Automation}} ({{ETFA}})}, pages 1--6, {Luxembourg, Luxembourg}, September 2015. {IEEE}.
\newblock ISBN 978-1-4673-7929-8.
\newblock \doi{10.1109/ETFA.2015.7301478}.

\bibitem[Loos et~al.(1996)Loos, Hokka, and J{\"o}rgl]{loosDynamicModelingCombustion1996}
H.~Loos, E.~Hokka, and H.P. J{\"o}rgl.
\newblock Dynamic {{Modeling}} of the {{Combustion Zone}} of a {{Waste Incineration Plant}}.
\newblock \emph{IFAC Proceedings Volumes}, 29\penalty0 (1):\penalty0 6897--6902, June 1996.
\newblock ISSN 14746670.
\newblock \doi{10.1016/S1474-6670(17)58791-1}.

\bibitem[Ljung(2023)]{ljungSystemIdentificationToolbox2023}
Lennart Ljung.
\newblock \emph{System {{Identification Toolbox User}}'s {{Guide}}}.
\newblock {MathWorks}, 2023a edition, 2023.

\bibitem[Ljung et~al.(2020)Ljung, Chen, and Mu]{ljungShiftParadigmSystem2020}
Lennart Ljung, Tianshi Chen, and Biqiang Mu.
\newblock A shift in paradigm for system identification.
\newblock \emph{International Journal of Control}, 93\penalty0 (2):\penalty0 173--180, February 2020.
\newblock ISSN 0020-7179.
\newblock \doi{10.1080/00207179.2019.1578407}.

\bibitem[Khosravi et~al.(2020)Khosravi, Iannelli, Yin, Parsi, and Smith]{khosraviRegularizedSystemIdentification2020}
Mohammad Khosravi, Andrea Iannelli, Mingzhou Yin, Anilkumar Parsi, and Roy~S. Smith.
\newblock Regularized {{System Identification}}: {{A Hierarchical Bayesian Approach}}.
\newblock \emph{IFAC-PapersOnLine}, 53\penalty0 (2):\penalty0 406--411, January 2020.
\newblock ISSN 2405-8963.
\newblock \doi{10.1016/j.ifacol.2020.12.200}.

\bibitem[Tacx et~al.(2021)Tacx, {de Rozario}, and Oomen]{tacxModelOrderSelection2021}
Paul Tacx, Robin {de Rozario}, and Tom Oomen.
\newblock Model {{Order Selection}} in {{Robust-Control-Relevant System Identification}}.
\newblock \emph{IFAC-PapersOnLine}, 54\penalty0 (7):\penalty0 1--6, January 2021.
\newblock ISSN 2405-8963.
\newblock \doi{10.1016/j.ifacol.2021.08.325}.

\bibitem[Sena et~al.(2021)Sena, Erkilinc, Dippon, Shariati, Emmerich, Fischer, and Freund]{senaBayesianOptimizationNonlinear2021}
Matheus Sena, M.~Sezer Erkilinc, Thomas Dippon, Behnam Shariati, Robert Emmerich, Johannes~Karl Fischer, and Ronald Freund.
\newblock Bayesian {{Optimization}} for {{Nonlinear System Identification}} and {{Pre-Distortion}} in {{Cognitive Transmitters}}.
\newblock \emph{Journal of Lightwave Technology}, 39\penalty0 (15):\penalty0 5008--5020, August 2021.
\newblock ISSN 1558-2213.
\newblock \doi{10.1109/JLT.2021.3083676}.

\bibitem[Kivits and {Van den Hof}(2019)]{kivitsDynamicNetworkApproach2019}
E.M.M.~Lizan Kivits and Paul~M.J. {Van den Hof}.
\newblock A dynamic network approach to identification of physical systems.
\newblock In \emph{2019 {{IEEE}} 58th {{Conference}} on {{Decision}} and {{Control}} ({{CDC}})}, pages 4533--4538, December 2019.
\newblock \doi{10.1109/CDC40024.2019.9030041}.

\bibitem[Ljung(2010)]{ljungPerspectivesSystemIdentification2010}
Lennart Ljung.
\newblock Perspectives on system identification.
\newblock \emph{Annual Reviews in Control}, 34\penalty0 (1):\penalty0 1--12, April 2010.
\newblock ISSN 13675788.
\newblock \doi{10.1016/j.arcontrol.2009.12.001}.

\bibitem[Arlot and Celisse(2010)]{arlotSurveyCrossvalidationProcedures2010}
Sylvain Arlot and Alain Celisse.
\newblock A survey of cross-validation procedures for model selection.
\newblock \emph{Statistics Surveys}, 4\penalty0 (none):\penalty0 40--79, January 2010.
\newblock ISSN 1935-7516.
\newblock \doi{10.1214/09-SS054}.

\bibitem[Brunzema et~al.(2022)Brunzema, Von~Rohr, and Trimpe]{brunzemaControllerTuningTimeVarying2022}
Paul Brunzema, Alexander Von~Rohr, and Sebastian Trimpe.
\newblock On {{Controller Tuning}} with {{Time-Varying Bayesian Optimization}}.
\newblock In \emph{2022 {{IEEE}} 61st {{Conference}} on {{Decision}} and {{Control}} ({{CDC}})}, pages 4046--4052, December 2022.
\newblock \doi{10.1109/CDC51059.2022.9992649}.

\bibitem[Garnett(2023)]{garnettBayesianOptimization2023}
Roman Garnett.
\newblock \emph{Bayesian {{Optimization}}}.
\newblock {Cambridge University Press}, March 2023.
\newblock ISBN 978-1-108-42578-0.

\bibitem[{The MathWorks Inc.}(2023)]{themathworksinc.StatisticsMachineLearning2023}
{The MathWorks Inc.}
\newblock \emph{Statistics and {{Machine Learning Toolbox User}}'s {{Guide}}}.
\newblock {MathWorks}, 2023a edition, 2023.

\bibitem[David et~al.(2023)David, Bernhardt, Beckmann, Krein, and Vodegel]{davidDeterminationFuelComposition2023}
Antje David, Daniel Bernhardt, Michael Beckmann, Anna Krein, and Stefan Vodegel.
\newblock Determination of the fuel composition during operation to optimise the combustion behaviour.
\newblock \emph{Fuel}, 343:\penalty0 127903, July 2023.
\newblock ISSN 00162361.
\newblock \doi{10.1016/j.fuel.2023.127903}.

\bibitem[Buekens(2013)]{buekensIncinerationTechnologies2013}
Alfons Buekens.
\newblock \emph{Incineration {{Technologies}}}.
\newblock {{SpringerBriefs}} in {{Applied Sciences}} and {{Technology}}. {Springer New York}, {New York, NY}, 2013.
\newblock ISBN 978-1-4614-5751-0 978-1-4614-5752-7.
\newblock \doi{10.1007/978-1-4614-5752-7}.

\bibitem[Beckmann et~al.(2005)Beckmann, Horeni, Metschke, Kr{\"u}ger, Papa, Englmaier, and Busch]{beckmannPossibilitiesProcessOptimization2005}
Michael Beckmann, Martin Horeni, J{\"o}rg Metschke, J{\"o}rg Kr{\"u}ger, Georg Papa, Ludwig Englmaier, and Michael Busch.
\newblock Possibilities of {{Process Optimization}} in {{Municipal Solid Waste Incineration Plants}} by an {{Online Balancing Program}}.
\newblock In \emph{Proceedings of the {{International Conference}} on {{Incineration}} and {{Thermal Treatment Technologies}}}, {Galveston, USA}, 2005.

\bibitem[Lips and Lens(2022)]{lipsDynamicModellingSimulation2022}
J.~Lips and H.~Lens.
\newblock Dynamic {{Modelling}} and {{Simulation}} of the {{Steam Cycle Utility System}} of a {{Waste Incineration Cogeneration Plant}}.
\newblock \emph{IFAC-PapersOnLine}, 55\penalty0 (20):\penalty0 139--144, January 2022.
\newblock ISSN 2405-8963.
\newblock \doi{10.1016/j.ifacol.2022.09.085}.

\end{thebibliography}

\end{document}